\newcommand{\logg}{$\log g\,$}        
\newcommand{\xfuv}{$m(1100)-m(1500)\,$}        %
\newcommand{\fnuv}{$m(1500)-m(2500)\,$}        %
\newcommand{\fuvv}{$m(1500)-V\,$}        %
\newcommand{\nuvv}{$m(2500)-V\,$}        %
\newcommand{\Teff}{$T_{\rm eff}\,$}        
\newcommand{\Tuv}{$T_{UV}\,$}        
\newcommand{\DYDZ}{$\Delta$$Y$$\!$/$\!$$\Delta$$Z\,$}        
\newcommand{\DM}{$\Delta M\,$}        
\newcommand{\Zsun}{$Z_\odot\,$}        
\newcommand{\Msun}{$M_\odot\,$}        
\newcommand{\Menv}{$M_{env}\,$}        
\documentstyle[11pt,aaspp4,psfig]{article}
\received{21 July 1997}
\slugcomment{accepted for publication in ApJ}

\lefthead{Yi et al.}
\righthead{The UV Upturn in Elliptical Galaxies}

\begin{document}

\title{On The Origin of the UV Upturn in Elliptical Galaxies. II. Test of the HB Hypothesis}
\author{Sukyoung Yi\altaffilmark{1,2} and Pierre Demarque}
\affil{Yale University, Department of Astronomy, P.O. Box 208101, New Haven, CT 06520-8101 \\ email: yi@shemesh.gsfc.nasa.gov demarque@astro.yale.edu}
\centerline{and}
\author{Augustus Oemler, Jr.}
\affil{Carnegie Observatories, 813 Santa Barbara St., Pasadena, CA 91101 \\ oemler@ociw.edu}

\altaffiltext{1}{present address: NASA/Goddard Space Flight Center, Code 681, Greenbelt, MD 20771}
\altaffiltext{2}{National Research Council Research Associate.} 

\begin{abstract}
   Whether or not metal-rich HB stars are the dominant UV source in giant
elliptical galaxies (gEs) is an important question in current astronomical 
research. 
   We follow up our previous evolutionary population synthesis study with 
quantitative tests to answer this question affirmatively under the following 
three conditions: (1) Reimers' empirical mass loss formula is proper, (2) the 
mass loss efficiency parameter ($\eta$) in metal-rich stars is somewhat larger 
than the value estimated from the metal-poor star studies, and (3) the true 
value of the helium enrichment parameter ($\Delta$$Y$$\!$/$\!$$\Delta$$Z$) 
is positive.
   All three important empirical characteristics of the UV upturn (i.e., the 
fact that strong UV upturns are restricted to gEs, the positive UV 
upturn-metallicity correlation, and the narrow range of the \Teff of the UV 
sources) are closely reproduced for reasonable ranges of input parameters.
   We discuss the major sources of uncertainties in the models, such as the 
production and role of hot horizontal-branch stars in gEs, and the importance 
of galactic nucleosynthesis.
\end{abstract}


\clearpage
\section{Introduction}

   The ultraviolet (UV) upturn phenomenon in the spectra of giant
elliptical galaxies (gE's) has been known since early space observations
with UV capability became available (\cite{cw79}).
   It is defined as the increase in flux with decreasing wavelength in the 
range $\approx$ 1,000 -- 2,500~\AA, as shown in Figure 1.

   Several important discoveries have been made related to the UV upturn.
Firstly, strong UV upturns are found only in the spectra of 
gEs\footnote{Metal-poor Galactic globular clusters show high ratios of
UV-to-$V$ flux (e.g., van Albada, de Boer, \& Dickens 1981) mainly because of 
opacity effects (Dorman, O'Connell, \& Rood 1995; Yi, Demarque, \& Oemler 
1997 - hereafter YDO). 
   However, their UV spectra are 
either flat or declining with decreasing wavelength, producing low ratios of 
far-UV-to-near-UV flux. Thus, it is correct to say that only gEs show a UV 
upturn with a steep slope in the UV spectrum.}.
   Secondly, IUE observations suggest a positive correlation between the 
magnitude of the UV upturn and Mg$_2$ index (\cite{f83}; \cite{b88}).
   If the Mg$_2$ index traces metallicity 
(although there is reason for caution [Worthey, Faber, \& Gonzalez 1992]), 
this implies that a more metal-rich galaxy shows a stronger UV upturn.
   Lastly, Hopkins Ultraviolet Telescope (HUT) observations suggest
that the sources of the UV photons are hot stars with a narrow range of 
temperature, i.e., \Teff $\approx$ 20,000 -- 23,000 K (Brown, Ferguson, 
\& Davidsen 1995).
   Since the dominant light sources (main sequence [MS], red giant branch 
[RGB], and horizontal branch [HB] stars) all tend to become cooler as 
metallicity increases, the unexpectedly high UV flux in such old, metal-rich 
systems has been a puzzle.

\placefigure{fig1}

   Understanding the cause of the UV upturn is important for the following 
reasons:
   (1) it provides insight into the hot stellar component in elliptical 
galaxies, (2) it tests the stellar evolution theory, (3) it constrains the 
age and metallicity of the majority of stars in gEs, if the UV upturn is 
sensitive to age and metallicity as some models suggest
(e.g., \cite{gr90}; Bressan, Chiosi, \& Fagotto 1994; \cite{dor95}; YDO).
   The age-dependence of the UV upturn is particularly noteworthy because
such models predict that the UV upturn becomes significant only at large ages 
when optical spectral evolution is hardly detectable. 
   Finally, (4) the UV upturn implies significant corrections to 
model-predicted optical
colors of distant (high redshift) galaxies (\cite{grv87}; \cite{bcf94}). 

   The origin of the UV upturn has been controversial since the first
observations were made, and several interpretations have been proposed.
   Young MS stars were among the favorite candidates as the UV sources 
in many studies (e.g., Gunn, Stryker, \& Tinsley 1981; \cite{grv87}; 
\cite{r88}; \cite{mb93}). 
   However, no evidence of recent star formation has been found in 
the UV-strong galaxies (\cite{o92}; \cite{ber93}). 
   Using the HUT, Ferguson et al. (1991) also found that a lack of C\,IV
absorption and the shape of the continuum were inconsistent with flux
from a MS population having a standard initial mass function. 
   Moreover, such hot MS stars (\Teff $\approx$~20,000 K: spectral type B)
are short-lived. 
   If the UV upturn were caused by young MS stars, it would be a 
transient feature, suggesting that all these UV-strong galaxies had 
experienced a secondary starburst recently, nearly at the same time, which 
is very unlikely.
   Post asymptotic giant branch (PAGB) stars were the next to attract 
attention (\cite{bc93}; \cite{mb93}).  
   However, PAGB stars are also thought to be so short-lived 
that the number needed to reproduce the UV upturn in the UV-strong gEs
would exceed that allowed by the fuel consumption theorem (\cite{ct91}).
   In addition, during most of their lifetimes, PAGB stars are much hotter 
than the suspected UV sources in gEs.
      
   Core helium-burning stars (HB and evolved HB stars) soon became an
attractive candidate because they also can be hot and bright (\cite{gr90},
and references therein).
   In addition, their mean temperature can match the estimated temperature of 
the dominant UV source in gEs easily and does not change 
rapidly with time, thus having advantages in explaining the narrow range of 
the \Teff of the UV sources.
   Since the HB in Galactic globular clusters tends to become hotter as 
metallicity decreases, {\it metal-poor} HB stars have been suggested as the 
cause of the UV upturn (\cite{aar78}; \cite{ay87}; \cite{l94}; \cite{pl97}).
   However, even the most metal-poor $and$ oldest Galactic globular clusters 
do not show UV upturns that are as strong as those in UV-strong 
gEs (Dorman et al. 1995; YDO).
   Moreover, gEs are metal-rich.
   Thus, if the metal-poor HB stars were the major UV sources in gEs, 
the positive UV upturn-metallicity relation would be puzzling, unless 
even the metal-rich gEs contain a substantial number of metal-poor stars $and$
the metal-poor stars in the UV-strong galaxies are significantly older than 
the oldest Galactic globular clusters (\cite{pl97}).

   Instead, Demarque \& Pinsonneault (1988) suggested that, under the
conventional assumptions of mass loss\footnote{Horch et al. (1992) proposed
that if mass loss on the RGB increases with metallicity, a more metal-rich
population would contain more hot (low-mass) HB stars. This assumption is
not empirically proven yet but consistent with Reimers' empirical
formula of mass loss for a fixed efficiency (see YDO).} and galactic helium 
enrichment, low-mass HB stars evolve into UV-bright objects instead of
becoming AGB stars.
   They found that this phenomenon, the so-called ``slow blue phase'' (SBP,
[Horch, Demarque, \& Pinsonneault 1992])\footnote
{The significance of the SBP is in its positive metallicity 
dependence. In some sense, the SPB phenomenon states the metallicity 
dependence of the combined AGB-manqu\'{e} (\cite{gr90}) and 
post-early-AGB (\cite{ct91}) evolutionary phases, a point 
which had not been addressed before. Yi, Demarque, \& Kim 
(1997, hereafter YDK) presented a mathematical analysis of the SBP 
and clarified the general confusion between the SBP and other terms.}, 
occurs more easily when metallicity is higher if $Z \gtrsim$ \Zsun.
   Then, the classical metallicity dependence of HB morphology (i.e., 
HB becomes redder as metallicity increases) should be reversed in the 
metal-rich regime ($Z \gtrsim$ \Zsun).
   Several evolutionary population synthesis (EPS) studies qualitatively 
showed that the hypothesis that such metal-rich, UV bright, core 
helium-burning stars are likely to be the dominant UV source in gEs 
is consistent with empirical data (\cite{gr90}; \cite{bcf94}; \cite{dor95};
\cite{yado95}; \cite{bfdd97}; YDO). 
   We call this {\it the metal-rich HB hypothesis}.

   In this quantitative study, we show, following YDO, that EPS models 
based on the metal-rich HB hypothesis reproduce quite well the empirical 
discoveries related to the UV upturn phenomenon.  
   We explore the sensitivity of the UV upturn in the models 
to the input parameters.  
   We compare single abundance models and a few composite models to 
observations of gE's.  
   We then discuss major uncertainties in the EPS models and the origin of
the discrepancies between various EPS studies.  
   Finally, the implications of the UV upturn for understanding galaxy 
evolution are also discussed.

\section{EPS Models}
   
   We have used the YDO model spectra (\cite{ydo97}) that were constructed for 
stellar systems of Age = 1 -- 25 Gyr with an age step of 1 Gyr, $Z$ = 0.0004, 
0.004, 0.01, 0.02, 0.04, 0.06, \& 0.1 and \DYDZ = 2 \& 3. 
   YDO synthesized the advanced evolutionary stages using Reimers' empirical
mass loss formula (\cite{r75}) with the mass loss efficiency parameter 
$\eta$ = 0.3, 0.5, 0.7, \& 1.0 (\cite{r81}) and truncated Gaussian mass 
distribution of the mass loss with mass dispersion factor 
$\sigma$ = 0.06~\Msun (see YDO for details).
   They assume that all stars in a gE formed in an instantaneous starburst.

   The EPS-sensitivity study of YDO concludes that the metal-rich HB 
hypothesis is plausible if $\eta \gtrsim$ 0.7 in metal-rich ($Z \gtrsim 0.01$) 
stars, assuming \DYDZ = 2 -- 3, $\sigma \approx$~0.06~\Msun, and a Salpeter
initial mass function (IMF). 
   YDO found that, other parameters being fixed, a positive
metallicity-dependence of $\eta$ can explain the UV-to-$V$ flux ratios of
both Galactic globular clusters and gEs.
   The dependence is approximately as follows: $\eta \approx$ 0.3 -- 0.5
($Z \lesssim 0.001$), $\eta \approx$ 0.5 -- 0.7 ($Z \approx 0.004$), 
and $\eta \gtrsim$ 0.7 ($Z \gtrsim 0.01$), while exact values depend on the 
adopted ages of the systems, i.e., as a smaller age is assumed, a larger 
$\eta$ is required.
   YDO call it {\it the variable-$\eta$ hypothesis} which is supported
by the independent theoretical work of the Iowa State group (\cite{bw91}; 
Willson, Bowen, \& Struck 1996).
   Therefore, in our single abundance models, we have adopted the YDO models 
of $\eta = 0.5$ for the models of $Z$ = 0.0004, those of $\eta = 0.7$ for 
$Z$ = 0.004 and 0.01, and those of $\eta = 1.0$ for the models of 
$Z \geq$ 0.02\footnote{The choices of $\eta$ in metal-poor stars are the
upper limits. Thus, we are overestimating the temperature of average HB stars 
in a system. However, since we are somewhat reluctant to perform 
interpolations between models of two different values of $\eta$,
we arbitrarily chose one value, which happened to be the upper limit, in our 
synthesis. The effect of such choices is small in the galaxy study because
metal-poor stars are minor sources of light in gEs.}.

   YDO also found that the Gaussian mass dispersion parameter, $\sigma$,
is important to the magnitude of the UV upturn, suggesting that a realistic 
synthetic HB construction must be employed in the UV population synthesis. 
   However, little is known about the true dispersion, and, thus,
we have adopted the models with a value of $\sigma$=~0.06~\Msun that was 
suggested by several globular cluster-HB morphology studies (e.g., Lee, 
Demarque, \& Zinn 1990, see YDO for details).
   YDO found that the sensitivity of an EPS model to $\sigma$ is moderate if 
$\sigma$ is uncertain only within a factor of 2, as suggested by Lee et al. 
(1990).   

   EPS studies (\cite{gr90}; YDO) suggested that the UV upturn is positively 
correlated with the helium enrichment parameter, \DYDZ, because both the 
stellar evolutionary pace and the UV bright, core helium-burning phase 
(the SBP) are sensitive to the helium abundance.
   Since the true value of \DYDZ is not well-known (see Section 6.2 for
discussion), we have carried out our study for two values, i.e., 
\DYDZ = 2 \& 3, for which complete evolutionary tracks are available. 

   YDO found that UV-to-$V$ flux ratios are not sensitive to the IMF slope.
   Thus, we have adopted the popular Salpeter (1955) IMF (with a slope of 
$-1.35$) in this study.

\section{Magnitude of the UV Upturn}

   The first task that EPS models must accomplish is to explain the magnitude 
of the UV upturn which seems to be positively correlated with metallicity.
   Why do only gEs exhibit strong UV upturns? Does the element that is 
responsible for the UV upturn in gEs also produce the UV upturn-metallicity 
relationship?
   The most obvious difference between gEs and other old stellar systems
(e.g., globular clusters and small elliptical galaxies), other than size and 
brightness, is metallicity: gEs are more metal-rich.
   According to YDO, metallicity plays a positive 
role in the magnitude of the UV upturn, and, thus, the metal-rich HB 
hypothesis seems qualitatively consistent with the empirical data.

   A quantitative comparison of the magnitude of the UV upturn between models 
and observations has been carried out. 
   The model magnitudes are defined as 
$m_\lambda$~$=$~$-2.5$~log~$<\!f_\lambda\!>$ 
where $<\!f_\lambda\!>$ is the mean flux in the bandpass. 
   The $<\!f(1100)\!>$, $<\!f(1500)\!>$, $<\!f(2500)\!>$, and $<\!f(V)\!>$ 
are defined by averaging the flux within the ranges 1,050 -- 1,200 \AA, 
1,250 -- 1,850 \AA, 2,200 -- 2,800 \AA, and 5,055 -- 5,945 \AA\, 
(\cite{a76}), respectively.
   Note that colors defined by such magnitudes are ratios of average 
fluxes rather than integrated colors over certain bandpasses.
   The observational data are from Table 2 of Dorman et al. (1995).
   We excluded two star-forming galaxies, NGC\,205 and a NGC\,4382, from their 
list.

   As YDO pointed out, \fuvv and \nuvv are not good measures of the strength 
of the ``UV upturn'' mainly because opacity effects are large in the near-UV.
   For instance, metal-poor Galactic globular clusters generally exhibit 
bluer UV-to-$V$ flux ratios than UV-strong gEs (\cite{dor95}; also Figure 18 
of YDO), even though they do not contain a substantial number of hot stars.
   A UV upturn, such as the ones shown in the spectra of gEs, can only be
generated by a strong bimodality in the temperature of stars 
(\cite{np85}; \cite{f95}), a characteristic
of metal-rich populations (see Figures 7-9 of YDO). 
   Opacity effects are less severe in \fnuv, and therefore \fnuv is 
mostly sensitive to the fraction of hot stars to cool stars.
   Thus we chose \fnuv as a UV upturn index.

\subsection{Single Abundance Models}

   It is useful to examine first the sensitivity of the UV upturn 
to age and to metallicity using single abundance models, before we construct
more elaborate models.
   Figure 2 shows the UV upturn index, \fnuv, as a function of age and 
metallicity. 
   The models shown in Figure 2 are based on the variable-$\eta$ hypothesis:
$\eta$ = 0.5 for $Z$ = 0.0004, $\eta$ = 0.7 for $Z$ = 0.004 and 0.01, and 
$\eta$ = 1.0 for $\geq$ 0.02, respectively (see Section 2 for the choice of 
$\eta$).    
   The observed flux ratios of gEs are marked as open circles at the left end 
of the left panels because their ages are unknown.

\placefigure{fig2}

   The average flux ratios of globular clusters are marked as open boxes with
a one standard deviation error bar (data are from Table 1 of Dorman et al. 
[1995]).
   The mean metallicity of the star clusters shown here is $Z \approx$ 0.0014 
(s.d. 0.0025).
   Figure 2 shows that the metal-poor models ($Z$ = 0.0004 -- 0.004) reasonably
match the cluster data at the age of 15 Gyr (the same stellar evolution 
models that have been used in this study yield 15 Gyr of age for a typical 
old cluster cluster, Chaboyer et al. [1996]).
   As pointed out earlier, globular clusters are in general bluer in 
\fuvv and in \nuvv than metal-rich models, mainly due to their low 
metallicities.

   Although single abundance models may be very unrealistic, they fit the 
empirical flux ratios reasonably and provide several important clues to the 
UV upturn phenomenon.
   Firstly, it is shown in Figure 2 that a large age ($>$ 8 Gyr) is essential 
to produce a significant amount of UV flux, unless the metallicity of the 
majority of stars in gEs is extremely large.
   This is because it takes time for a population to develop a substantial
number of low mass core-helium burning stars.
   However, under the current assumptions, the empirical ranges indicate an 
average age of 10 -- 14 Gyr if \DYDZ = 2 for gEs, as shown as shaded boxes.
   This age estimate is somewhat smaller than the typical ages of the Galactic
globular clusters.
   Secondly, at a fixed age (e.g., 12 Gyr if \DYDZ = 2), more metal-rich 
models ($Z \approx$ 0.04, thick dotted lines) show stronger UV flux ratios 
than less metal-rich ones ($Z \approx$ 0.02, thick solid lines).
   This is because (1) more metal-rich stars lose more mass on the RGB 
according to the Reimers' mass loss formula and (2) a higher helium
abundance in a more metal-rich star (assuming a positive 
$\Delta$$Y$$\!$/$\!$$\Delta$$Z$) causes the SBP more prominent.
   If the majority of stars in gEs are metal-rich ($Z \gtrsim$ 0.02), as 
various studies suggest (Buzzoni, Gariboldi, \& Mantegazza 1992, and 
references therein), such a metallicity difference naturally explains the 
empirical UV upturn-metallicity relation.
   As YDO pointed out, this is because both the estimated stellar mass loss 
predicted by Reimers' formula and the SBP (slow blue phase) are positively
correlated with metallicity.
   Thirdly, the observed range of \fnuv, the UV upturn index, is easily
achievable only by metal-rich models, as shown in the bottom panels of 
Figure 2.
   Metal-poor models cannot reproduce the observed range of \fnuv in gEs 
unless a very large age ($\gg$ 20 Gyr) is adopted. 

   As an alternative to the metallicity spread, an age spread among gEs can
also produce the observed difference in UV flux ratios if UV-strong gEs are
older than the weak ones by 10 -- 30\%.
   A metallicity spread, rather than an age spread, would be a more natural 
choice if the empirical UV upturn-metallicity relation is real.
   Conversely, if there is no direct causality between the strength of the 
UV upturn and metallicity (see the discussion of Lee [1994]), the study
of the UV upturn may constrain the relative ages among gEs!

\subsection{Composite Models with Realistic Metallicity Distributions}
   
   It is certainly an oversimplification to assume that all the stars in 
a gE have the same metallicity (\cite{l74}; \cite{ya87}; \cite{g97}). 
   Thus, we investigate if any physically plausible mixture of different 
abundance models can match the data.

   The chemical evolution of gEs is still poorly understood. 
   For example, neither the true value of \DYDZ, nor the metallicity 
distribution of stars in any galaxy is yet clear.
   We do not quite understand how gEs have achieved the high metallicities 
that are measured (e.g., \cite{ari97}), either.
   Popular theories suggest an instantaneous chemical enrichment at an early 
stage of galactic evolution, converting most of the gas into stars, in order 
to explain the high metallicities of gEs (\cite{l74}; \cite{tl79}; \cite{ya87},
\cite{wpm95}; \cite{ari97}; \cite{g97}).
   But, it is still unclear why gEs had to go through such explosive 
starbursts at the same epoch. 
   While we admit our ignorance on the galactic chemical evolution, we have
adopted the popular abundance distribution models of Yoshii \& Arimoto (1987, 
hereafter YA87) and of Tantalo et al. (1996, hereafter ``Infall'' models).

   Figure 3 displays four composite models based on the YA87 distributions. 
   The building blocks are the single abundance models  described in 
Section 3.1.
   Models shown as solid lines are based on the YA87 $10^{12}$\,\Msun model, 
the majority of whose stars are $Z \approx$ 2~\Zsun.
   The other models (dashed lines) are for a galaxy of 
$4\,\times\,10^{9}$\,\Msun in which the majority of stars have 
$Z \approx$~\Zsun.
   These two models are believed to approximately cover the metallicity range 
of gEs whose optical spectra suggest a high metallicity 
($\gtrsim$~\Zsun)\footnote{We do not argue that these masses defined by 
the YA87 models are adequate for gEs. We only tried to adopt realistic 
metallicity distributions whose average metallicities match those of gEs.}. 

\placefigure{fig3}

   YDO suggested that $\eta$ in metal-rich stars should be larger than 0.7
for the metal-rich HB hypothesis to work.
   In order to indirectly estimate the true $\eta$ in metal-rich stars,
we have constructed models that are based on the variable-$\eta$ 
hypothesis, but with a different $\eta$.
   The thin lines are the models with the $\eta$ cutoff at $\eta_{max}$ = 0.7;
i.e., $\eta$ = 0.5 for $Z$ = 0.0004 and $\eta$ = 0.7 for $Z \geq$ 0.004.
   These models generate the lower bounds of the predicted UV flux.
   Meanwhile, the hydrodynamical simulations of the Iowa State 
group\footnote{Iowa State group's simulations were carried 
out only up to $Z$ = \Zsun. We are not certain which value of $\eta$ such a 
study would predict for the stars of $Z >$ \Zsun. So, we used the same 
$\eta$ suggested for $Z$ = \Zsun (i.e., $\eta = 1.0$) for the stars of 
$Z \geq$ \Zsun.} suggest $\eta$ = 1.0 for $Z$ = \Zsun.
   Thus, we have constructed another set of models with $\eta_{max}$ = 1.0;
i.e., $\eta$ = 0.5 for $Z$ = 0.0004, $\eta$ = 0.7 for $Z \geq$ = 0.004 \& 0.01,
and $\eta$ = 1.0 for $Z \geq$ = 0.02.

   Interestingly, composite models based on the YA87 prescription deviate
from the data in \nuvv significantly, by predicting too much near-UV flux.
   Since much of the near-UV flux in the composite model comes from metal-poor 
stars, this may indicate that the YA87 models predict too many metal-poor 
stars.
   This is analogous to the G-dwarf problem in our Galaxy (\cite{at76}), and
it has been noticed by others (\cite{tan96}; Worthey, Dorman, \& Jones 1996).
   Tantalo et al. (1996) claimed that this high near-UV-flux problem can
be alleviated if the chemical evolution model takes into account gas that
falls in.
   Such infall models have been known to produce fewer metal-poor stars
(\cite{l72a}; \cite{l72b}; \cite{at76}; \cite{c80}; \cite{tan96}; \cite{k97}) 
than non-infall type models, such as the YA87 models.
   In addition, infall models have a lower maximum metallicity
(e.g., $Z_{max} \approx$ 0.07 in Tantalo et al.'s infall models and 0.08
in YA87 models, respectively.).

   Figure 4 shows the composite models based on the abundance distributions 
produced by the infall model. 
   The thin (thick) lines are based on $\eta_{max}$ = 0.7 (1.0).
   Solid (dashed) lines are based on the model distributions of $10^{12}$ 
($5 \times 10^{11}$) ~\Msun and their mean metallicities are approximately
2\,\Zsun (\Zsun), respectively.
   Due to the smaller fraction of metal-poor stars in the infall model,
the composite models of $\eta_{max}$ = 0.7 -- 1.0 now consistently match the 
empirical data both in the far-UV and in the near-UV.
   
\placefigure{fig4}

   Note that models of different \DYDZ lead us to different interpretations 
about the empirical relation between the strength of the UV upturn and 
metallicity.
   The \DYDZ = 2 models predict little difference in UV flux ratios as
metallicity differs by a factor of two.
   In this case, the observed ranges in flux ratios are likely caused by
an age difference among gEs, in the sense that UV-stronger galaxies are older
by about 10 -- 30\%.
   Then, the empirical UV upturn-metallicity relation has to be either 
spurious or a consequence of other indirectly-connected relations.
   In contrast, the \DYDZ = 3 models suggest that a factor of two difference
in metallicity is sufficiently large to cause the UV upturn-metallicity
relation at a given age.
   This is because the metallicity-dependence of the SBP (slow blue phase) 
of core helium-burning stars, the dominant UV source, is more significant 
when the helium abundance is larger.
   Similarly, an age difference of about 10 -- 30\% among gEs can also produce
an equivalent amount of difference in the UV strength.
   More reliable metallicity measurements would remove such entanglements 
between age and metallicity.
 
   Let us see whether such composite models, whose flux ratios are in 
reasonable agreement with data, match the overall spectral shape.
   Figure 5 exhibits a 13 Gyr old infall model and the empirical spectrum
of NGC\,4552\footnote{
   This model is not the best-fitting model but a model whose flux ratios are
in reasonable agreement with those of NGC\,4552.
   The empirical spectrum is not entirely that of NGC\,4552, as described in 
the figure caption of Figure 1.
   Thus, we are not sure if this kind of mosaic spectrum is reliable enough to
be used in detailed fitting.
   Unfortunately, we were not able to find a galaxy whose well-calibrated 
spectrum is available from far-UV to infrared.}.
   They have been normalized in the $V$ band.
   The overall match is good.
   However, even though the infall model has fewer metal-poor
stars compared to the non-infall type YA87 model, the model still seems to
overproduce the flux in the range of 1,700 -- 3,600 \AA\, by as much as 20 -- 
50\% (c.f., \cite{do96}). 
  Could this be still caused by uncertainties in the metallicity distribution? 

\placefigure{fig5}

   Figure 6 shows that, when a model approximately fits the spectrum of 
NGC\,4552 (middle panel), the dominant UV source is metal-rich.
   However, in the 1,700 -- 3,600 \AA\, range, the light contribution from 
metal-poor stars no longer dominates.
   It seems that the problem of the overpredicted near-UV flux would not 
disappear entirely, even if we remove the whole metal-poor stars from the 
model.
   It is unlikely that the discrepancy is caused entirely by the uncertainties 
in the fraction of the metal-poor stars in the model metallicity distribution.

\placefigure{fig6}

   Figure 7 shows the light contribution from less evolved (MS \& RGB) stars 
and from evolved (post-RGB) stars. 
   Post-RGB stars make up most of the UV flux and about 20\% of optical flux. 
   The flux from post-RGB stars have two peaks, one in the far-UV and
the other approximately in the $V$ band.
   The far-UV peak is mostly caused by hot HB and evolved HB stars,
whereas the optical peak is caused mostly by AGB and red HB stars.
   The flux from such evolved stars is quite uncertain, mainly because the
mass loss both on the RGB and on the AGB is poorly understood.
   It seems that at least the overestimated flux in the range 1,700 --
2,500 \AA\, is related with the predicted HB mass distribution, because,
in this wavelength range, MS and RGB stars contribute little.
   If the true flux near 2,500 \AA\,  from the evolved stars is lower by about 
50\%, the problem with the overestimated model flux will disappear.
   One way of producing a lower flux in this valley is to have a temperature 
distribution of HB stars that is even more strongly bimodal than our models 
predict.
   We will discuss this question later in Section 6.

\placefigure{fig7}

   Despite such uncertainties, composite models, mainly based on the Infall 
prescription, are successful in matching the observed flux ratios in various 
bandpasses.
   As pointed out in the previous section, more metal-rich models show 
stronger UV upturns, if metallicity differs by a factor of two. 
   This trend agrees with the empirical data.
   On the other hand, if no significant metallicity difference exists among 
gEs, a moderate age difference between UV-strong and UV-weak galaxies 
seems to be required in the sense that UV-strong galaxies are older than the 
UV-weak ones by 10 -- 30\%.
   Whether gEs have such a substantial age difference among themselves is 
still an open question (c.f., Faber, Worthey, \& Gonzalez 1992; \cite{ka97}).

\subsection{Regarding ``The Onset of the UV Upturn''}

   The different maximum metallicity predicted by infall models and non-infall 
type models has a profound impact on the predicted timing of ``the onset of 
the UV upturn''.
   For example, the Padova group (\cite{bcf94}; \cite{tan96}) found that their 
models with \DYDZ = 2.5 experience the onset of the UV upturn at 5.6 Gyr.
   Our models (Figures 3 \& 4) confirm their suggestion qualitatively.
   This is potentially an important discovery, but, one has to be careful
in applying this concept.
   Let us examine this more closely.

   The large bump near 5 Gyr (3 Gyr) of age in the \DYDZ = 2 (3) models based 
on the YA87 $10^{12}$~\Msun model (top and bottom in Figure 3) is caused 
mostly by the UV light coming from extremely metal-rich ($Z >$ 0.06) stars.
   Such metal-rich stars are also helium-rich under the current assumption 
of a positive \DYDZ, and, in fact, it is the helium abundance that is supposed
to be more directly related with the UV light production (\cite{dro93}; YDO).
   For this reason, the onset of the UV upturn occurs earlier if a larger 
\DYDZ is adopted, as shown in Figures 3 \& 4.

   These extremely metal-rich stars produce a tremendous amount of UV light at 
small ages, as shown in Figure 2 (and also in Figures 9 \& 12 of YDO).
   However, the most metal-rich group of stars fade away early because they 
evolve faster than less metal-rich stars when a positive \DYDZ is assumed.
   Then, assuming a smooth metallicity distribution (although our EPS models
use discrete distributions), the next metallicity group begins to produce
UV light, and so on (see Figure 2).
   This is perhaps the manner in which a gE develops a strong UV upturn.
   Thus, the timing of the onset of the UV upturn is sensitive to the maximum 
metallicity.
   Note that such a sequential development of UV flux is not predicted
by single abundance models (Figure 2).

   Since the infall models have a smaller maximum metallicity than YA87 models 
do, the onset of the UV upturn with the infall models is later, 
as shown in Figures 3 \& 4.
   Tantalo et al.'s estimate of the onset of the UV upturn, 5.6 Gyr, seems to
originate from the extremely metal-rich ($Z$ = 0.1) stars whose characteristic
onset of the UV light production occurs at about 5.6 Gyrs, as shown in
their Figure 4.
   Since the maximum metallicity in their infall models is $Z \approx$ 0.07 
(see their Figure 12) and their building blocks have a metallicity grid of 
$Z$ = 0.0004, 0.004, 0.008, 0.02, 0.05, and 0.1, they seem to have used the 
$Z$ = 0.1 population to represent the extremely metal-rich ($Z >$ 0.05) stars 
that may not be represented well by their stellar models of $Z$ = 0.05.
   We can show the effect of the metallicity grid as well.
   In our Figure 4, we used Tantalo et al.'s infall models.
   Since our building blocks have different metallicity grids, i.e., 
$Z$ = 0.0004, 0.004, 0.01, 0.02, 0.04, 0.06, \& 0.1, we used the 
$Z$ = 0.06 population, instead of $Z$ = 0.1, to represent the stars of 
$Z$ = 0.05 -- 0.07.
   Then, the onset of the UV upturn is delayed significantly (to about 10 Gyr 
\& 6 Gyr in the case of \DYDZ = 2 \& 3, respectively, Figure 4).
   If we use  \DYDZ = 2.5 as Tantalo et al. did, our models would suggest
approximately 8 Gyr as the onset of the UV upturn.
   So, we argue that their estimate, 5.6 Gyr, is a product of their choice
of metallicities for building blocks.
   But, the basic concept still holds good qualitatively.

   In principle, a study (including observations) of the onset of the UV 
upturn would put a useful constraint on the metallicity distribution in a gE. 
   However, in practice, the onset of the UV upturn is still quite uncertain 
because it is also sensitive to any input parameter that is influential to the 
production of evolved stars, such as \DYDZ, $\eta$, metallicity distribution,
and even $\sigma$ (the dispersion on the mass loss).

\section{Characteristic Temperature of the UV Upturn}

   What are the observable properties of the stars that are responsible for the
UV upturn? Do our models reproduce not only the magnitude of the UV upturn,
but also the properties of the UV sources? 
   It is extremely difficult to determine the physical properties 
of the major UV sources, such as mass and luminosity, from the integrated
spectrum. 
   But, it is possible to determine the probable effective temperature of the 
dominant UV sources reasonably well if their representative metallicity can be
constrained independently. 
   This is because the shape of the UV flux is governed mostly by the 
temperatures of the UV sources.

   To begin with, it is convenient to define the characteristic temperature 
of the UV upturn, \Tuv, as the effective temperature of the star whose 
spectrum is in best agreement with the UV spectrum of the galaxy.
   Precise determination of \Tuv is important because it tests the
validity of the stellar evolution theory and accompanying assumptions
in the EPS. 
   Any successful population synthesis model should reproduce not only the 
magnitude of the UV upturn but also the observed \Tuv.
   However, since not all UV sources have the same temperature, the \Tuv
determination is not trivial. 
   It is more difficult when \Tuv is to be estimated from the spectrum whose 
UV upturn is not strong, because, then, the UV light contribution 
from less hot stars, such as MS stars, becomes important.
   
   The flux ratio, \xfuv, is 
a reasonable \Tuv indicator, because the slope of the spectrum in the range 
$\lambda$ = 1,000 -- 1,500 \AA\, is sensitive to \Teff.
   Figure 8 shows \xfuv as a function of \Tuv and metallicity. 
   The value, \xfuv, is certainly affected by metallicity because of the 
severe line-blanketing effect in the UV. 
   Although not shown in this plot, \xfuv reaches a maximum
near \Teff = 50,000 (\xfuv $\approx -1.0$), then it drops to $- 0.6$ or so.
   For example, stars of \Teff = 25,000~K and 60,000~K would have similar 
values of \xfuv.
   Thus, \xfuv alone does not provide a unique \Tuv.
   However, their spectra look very different below the $Lyman$ break, and,
thus, one can easily distinguish one from another by looking at the far-UV
spectrum.
   Therefore, in practice, \xfuv serves as a reliable \Tuv indicator.  
   The empirical values of \xfuv of seven gEs have been measured from
the HUT spectra (\cite{bfd95}) and marked as open circles in Figure 8.
   The data have been placed at the left end of the figure because their
\Tuv's are not known a priori.
   If the dominant UV sources in gEs are metal-rich (e.g., $Z \gtrsim$~\Zsun), 
as the HUT team suggested (\cite{bfdd97}), Figure 8 tells us that the 
empirical \Tuv is approximately 19,500 -- 23,400~K, which is in good agreement 
with Brown et al.'s (1995) estimate, \Tuv $\approx$ 20,000 -- 23,000~K. 

\placefigure{fig8}

   The UV-strong gE, NGC\,1399, also has a lower value of \xfuv, suggesting 
that the UV source in this galaxy is hotter than those in other observed gEs.
   Note that, however, the two other UV-strong gEs, NGC\,4552 and 
NGC\,4649 (categorized in the UV-strongest galaxy group by Burstein et al. 
1988), show larger values of \xfuv (lower values of \Tuv) than the other
UV-weak gEs, indicating that the dominant UV sources in these 
UV-strong gEs are cooler than those in UV-weak gEs.
   This trend has also been noticed by Brown et al. (1997, see their Fig. 7).
   We wonder if such correlation between \Tuv and the magnitude of the 
UV upturn is real and why NGC\,1399 is an exception.
   If a larger sample confirms this, models certainly should explain it.
   We attempt to explain this discovery in this section.

\subsection{Single Abundance Models}

   Single abundance models have been constructed and compared with empirical
data in Figure 9. 
   The narrow shaded box is the observed range of \xfuv defined by the
seven HUT spectra whose estimated values of \xfuv are shown in Figure 8, 
assuming these galaxies are between 10 and 20 Gyrs old. 
   The models were constructed with the same parameters as shown in Figure 2.
   As in the case of the magnitude of the UV upturn, \DYDZ affects the model 
predictions only in terms of the required age.

\placefigure{fig9}

   Metal-rich models show a constant high flux ratio (\xfuv $\approx -0.6$) 
at low ages (left of arrows (a) \& (b)).
   The arrows (a) \& (b) indicate the turning points of the $Z$ = 0.04 \& 0.02
model in age at which core helium-burning stars become more important UV 
sources than PAGB stars.
   In the PAGB epoch, \xfuv does not vary much.
   This is because PAGB stars, the dominant UV sources at small ages, follow
very similar evolutionary tracks as the masses of their progenitors vary,
according to conventional PAGB evolution models.

   Almost no metallicity dependence is visible between the models of $Z$ = 0.02
and 0.04.
   This may be largely an artifact because YDO constructed their EPS models 
using the same PAGB tracks (tracks of the Kiel group: \cite{s79}; \cite{s83}; 
\cite{bs90}) for different metallicities and using the same stellar spectra 
for stars of \Teff $>$ 50,000~K.
   But, the effect of metallicity on the stellar spectrum should not be large
at high effective temperature.
   Unless PAGB tracks are very sensitive to metallicity (c.f., \cite{vw94}), 
this approximation should be reasonable.
   Despite such uncertainties, model flux ratios in \xfuv are in surprising 
agreement with the empirical values of UV-weak gEs (top end of the shaded box).
   If this constant \xfuv is not  an artifact, this would suggest 
that the dominant UV sources in UV-weak gEs are PAGB stars.
   This argument is consistent with the HUT team's conclusion on the dominant
UV source in the bulge of M\,31 whose UV upturn feature is weaker than those in
UV-strong gEs (\cite{fd93}).

   As a metal-rich galaxy ages, a larger number of hot (but, still cooler 
than PAGB stars) core helium-burning stars are produced, dominating over PAGB
stars in the far-UV and causing an increase in the value of \xfuv.
   Suppose that gEs have ages between the two turning points (a) and (b), 
that is, 10 -- 11 Gyr if \DYDZ = 2 (or 7.5 -- 9 Gyr if \DYDZ = 3), and that 
gEs are reasonably represented by metal-rich single-abundance models.
   Then, the puzzling phenomenon of UV-weak gEs showing a higher \Tuv 
(see the header of Section 4) can be explained by a metallicity 
difference, in the sense that UV-strong galaxies are more metal-rich by a 
factor of two in $Z$.
   Although this explanation is consistent with the explanation of the
UV upturn-metallicity relation, we are uncomfortable with such a fine
tuning of the age and metallicity of gEs.
   Alternatively, the higher \Tuv of UV-weak gEs can be explained if
UV-weak gEs are somewhat younger.

   Once core helium-burning stars dominate over PAGB stars as UV sources,
\xfuv decreases (becoming bluer) with increasing age, experiencing another
turning point, (c) for $Z$ = 0.04 and (d) for $Z$ = 0.02.
   From this second turning point on, core helium-burning stars dominate the 
UV flux, and as core helium-burning stars become hotter, \xfuv decreases.
   The UV-strong gE with a higher \Tuv, NGC\,1399, can be explained
if it is already passed the second turning point.
   Then, this would suggest either that NGC\,1399 is older than other gEs 
observed if no significant metallicity difference exists, or that it
is more metal-rich in which case the second turning point occurs earlier
if these gEs are of similar ages, or perhaps both combined.
   Since NGC\,1399 is one of the most metal-rich gEs (at least with a high 
Mg$_2$, [Dorman et al. 1995]), such explanations are reasonable.
   It is clear that, if metallicity is uncertain, \xfuv alone cannot 
constrain the age of the galaxy.

   Metal-poor models take quite different paths.
   At small ages, metal-poor models still have significant UV light 
coming from MS stars.
   Then, before MS stars become cool enough, hot HB stars develop and 
PAGB stars never have a chance to dominate the UV spectrum.
   Thus, metal-poor models never reach such a blue \xfuv epoch as metal-rich
models easily do, unless a very large age ($>$ 20 Gyr) is adopted.
   This shows that metal-poor models cannot match the empirical \Tuv.

\subsection{Composite Models}
   
   Unlike the moderate success of the single abundance models in matching the 
observed range of \xfuv, composite models (Figure 10), constructed in the same 
manner as those in Figure 4 using the infall models, do not match the observed 
range of \xfuv well.
   Models are slightly redder than observed.
   At least part of this is caused by the overestimated near-UV light 
discussed in Section 3.2. 
   If we trust our single abundance models, this would suggest that the 
mean metallicity of the majority of stars in a gE must be peaked around
approximately 1 -- 2 \Zsun.
   As the metallicity dispersion is larger, the \xfuv fits would become
worse.
   However, it would be premature to conclude so until single abundance models 
are empirically checked first.

\placefigure{fig10}

\section{Two Color Diagrams in the UV}

   Two color diagrams are presented in Figures 11 -- 12. 
   In Figure 11, empirical data (filled circles) are compared with single 
abundance models.
   The 13 Gyr models are marked as open circles with metallicity shown
next to them.
   Models of approximately solar abundance match the data reasonably if gEs
are about 13 Gyrs old.
   In the metal-rich regime, the effect of metallicity and that of age are 
almost on top of each other, making these quantities hard to extract from flux
ratios alone.
   However, for metal-poor populations, such two color diagrams based on 
UV flux ratios can provide a good age estimate.

\placefigure{fig11}
\placefigure{fig12}

   As pointed out earlier, composite models (Figure 12) somehow overpredict
the flux in the near-UV, even if infall models are adopted.
   At the moment, it is unclear whether this mismatch is caused by the
uncertainties in the chemical evolution model or by others, such as
those in the HB mass distribution treatment.

\section{Discussion: Uncertainties in the Models}

   It has been pointed out earlier that $optical$ colors derived from 
EPS models are sensitive to the uncertainties in the stellar evolutionary 
tracks and in the stellar spectral libraries 
(\cite{w94}; Charlot, Worthey, \& Bressan 1996).
   In the UV, however, so little is known about the late phase of stellar
evolution that even such basic input assumptions as mass loss and \DYDZ play 
significant roles (\cite{gr90}; \cite{jt93b}; YDO).
   YDO investigated the effects of mass loss, \DYDZ, the IMF slope $x$, and 
mass dispersion on the HB, among many sources of uncertainties.
   A few additional comments are discussed in this section.

\subsection{Mass Loss}

   Mass loss is one of the most influential input parameters to the UV flux
because it determines the masses (thus \Teff) of HB stars that
are important UV sources.
   However, little is known about the processes of mass loss.
   Reimers' empirical formula (\cite{r75}; \cite{r81}) has been widely 
used in many EPS studies, thanks to its simple parameterized form, but, the 
mass loss efficiency parameter, $\eta$, is not well-determined.
   Various population studies regard $\eta$ or the amount of mass loss as
a free parameter.
   Thus one should be aware of the effects of the assumed mass loss treatment
in order to decipher the results from a UV EPS study properly.

   For instance, if a very small $\eta$ ($<$ 0.5) is used, models produce a 
much lower UV flux at a given time{\footnote{
Somehow, the Padova group (\cite{bcf94}; \cite{tan96}) achieved an equivalent 
amount of UV flux to ours at a fixed age, even though they used a much lower 
$\eta$ (= 0.45) than ours (= 0.7 -- 1.0 for metal-rich stars).
   Since the UV flux is also sensitive to the amount and the dispersion of 
the mass loss, it would be necessary to compare such quantities in their 
studies to those in YDO.}.
   Such low $\eta$'s mostly stem from HB morphology fittings (see YDO) that 
are carried out for metal-poor stars.
   However, there seems to be a metallicity dependence of $\eta$; for example,
the UV spectra of relatively metal-rich Galactic globular clusters already 
indicate $\eta \approx$ 0.5 -- 0.7.

   For another example, Park (1995) and Park \& Lee (1996) used a fixed mass 
loss, that is \DM = 0.223~\Msun regardless of age and metallicity, in their 
EPS studies.
   This value, \DM = 0.223~\Msun, in fact corresponds to $\eta \approx$ 1.0
for $Z$ = 0.0001, $\eta \approx$ 0.7 for $Z$ = 0.004, $\eta \approx$ 0.5 for 
$Z$ = 0.02, and $\eta \approx$ 0.4 for $Z$ = 0.06 at the age of 15 Gyr
(see Figures 3 -- 4 in YDO), which is exactly opposite to the variable-$\eta$ 
hypothesis discussed in YDO.
   Remember that the metallicity dependence of mass loss for a fixed $\eta$
was one of the necessary conditions in producing a UV upturn (see Section 5).
   Thus this constant \DM assumption drives metal-poor populations in a 
composite galaxy model to produce more UV flux than the conventional $\eta$ 
($\approx$ 0.3 -- 0.5 in metal-poor stars) allows, while it suppresses 
metal-rich populations from developing hot HB stars and thus from generating 
a high UV flux.
   This is an extreme prescription, given present knowledge about 
mass loss.
   Consequently, they conclude that the UV upturn phenomenon in gEs is caused 
by metal-poor HB stars.
   This example clearly illustrates the high sensitivity of the UV population 
synthesis to the mass loss treatment.

   There has been a different approach to the mass loss, which
allows a dispersion in $\eta$ in red giants instead of assuming a dispersion
in \DM (\cite{jt93a}; \cite{ddro96}, see also \cite{cc93}).
   This is an interesting idea and may be plausible if some dispersions
in physical or chemical properties of red giants, such as rotation or
convection, are causing a difference in the mass loss efficiency (or $\eta$) 
which is a more fundamental quantity than \DM.
   However, a current model of this kind (\cite{ddro96}) seems to produce UV 
sources that are too hot (\Teff $\approx$ 28,000 K) to be consistent with the 
estimated temperature of the major UV sources in gEs (20,000 -- 23,000 K).
   This approach deserves attention, and its physical basis and impact 
on EPS models should be investigated further.

\subsection{\DYDZ}

   A fundamental basis of the metal-rich HB hypothesis is a postive 
$\Delta$$Y$$\!$/$\!$$\Delta$$Z$.
   A higher \DYDZ causes a model galaxy to develop a UV upturn more quickly
because of the evolutionary pace on the RGB and the SBP (slow blue phase)
phenomenon, both of which are positively correlated with helium abundance 
(YDK).
   What if \DYDZ $\ll$ 2? Then, the SBP phenomenon would be insignificant.
   J\mbox{\o}rgensen and Thejll (1993b) claimed that the metallicity 
dependence of the magnitude of the UV upturn can be achieved only when the 
true value of \DYDZ in gEs is larger than the value in the Sun.
   This is true if $\eta$ in metal-rich stars is significantly smaller than 
1.0.
   For instance, if we show the metal-rich single abundance models of 
$\eta$ = 0.3 -- 0.5 as a function of age and metallicity as in Figure 2,
our models would be consistent with what J\mbox{\o}rgensen and Thejll 
suggested.
   On the other hand, our models of $\eta$ = 1.0 predict a clear metallicity
dependence regardless of \DYDZ, as shown in Figure 2.
   We agree with J\mbox{\o}rgensen and Thejll in that the metallicity
dependence is more prominent as \DYDZ increases (see Figure 2).
   However, we find it very difficult to make a general statement about such 
trends because such trends are quite sensitive to the adopted parameters.
   It is more so if such crucial input parameters as $\eta$ depend on
metallicity.

   Since approximately one half of the UV light in old, metal-rich models is 
attributed to SBP stars, as shown in Figures 14 \& 15 in YDO,  a model galaxy 
without a UV light contribution from SBP stars would not be able to achieve 
such a prominent UV upturn in a reasonable time scale as observed in gEs. 
   In this sense, a moderate \DYDZ ($\gtrsim$ 2) is preferred to explain
the empirical data about the UV upturn, unless other parameters, such as $Z$ 
or $\eta$, are markedly different from the assumptions used in this work.
   The danger is that the current choice of \DYDZ  basically comes
from an extrapolation based on two points, that is, the assumed primordial
chemical composition ([$Z$, $Y$] $\approx$ [0.00, 0.23]) and the solar
composition ($\approx$ [0.02, 0.28]) (c.f., Dorman et al. [1995]).

\subsection{Mass Distribution on the HB}
   
   YDO showed that the UV flux is sensitive to the assumed mass distribution 
on the HB.
   A truncated-Gaussian function as an approximation to the HB mass 
distribution in Galactic globular clusters (\cite{r73}; \cite{ldz90};
\cite{ldz94}) has been questioned mainly because of two empirical findings:
   (1) The HB morphology of some Galactic globular clusters, e.g., M15, is 
not reproducible by a Gaussian function (\cite{r90}). 
   There are too many extremely hot HB stars with a thin envelope
(\Menv $\approx$ 0.05 -- 0.15~\Msun) to be explained by single-Gaussian models.
   (2) There are several hot stars in the old open cluster NGC\,6791 and for 
three of them the best fit is acquired by the evolved HB models of 
\Menv $\approx$ 0.005~\Msun (\cite{lsg94}).
   The presence of subdwarf B stars raises the same question (\cite{smd74}).
   These stars are not predicted to exist by the minimum \Menv hypothesis
that was discussed in detail by YDO and used in this study.
   The minimum \Menv hypothesis suggests that HB stars cannot have smaller
envelope mass than a critical value because their precursors, red giants, 
cannot initiate helium core flash at the tip of the RGB if \Menv is smaller
than the critical value and the helium core flash is not believed to remove
any significant amount of envelope material into space.
   While the exact value is still uncertain, YDO adopted 0.01~\Msun as the
minimum \Menv on the HB.
   See YDO for the detailed discussion that includes different views.
   As the minimum \Menv in the EPS decreases, the resulting
UV flux becomes stronger and \Tuv becomes higher.
   It is interesting to note that, if we discard the concept of the minimum 
\Menv and assume HB stars can have a virtually zero \Menv, then the problem 
of the low \Tuv in the composite models (Figure 10) may disappear.
   A further investigation is necessary.
   
   Is the single-mass Gaussian assumption appropriate? 
   We pointed out in Section 3.2 that our models produce too much near-UV flux.
   This could be a sign of the inadequacy of a single-mass Gaussian function 
for the HB mass distribution.
   If the true HB temperature distribution is somehow more bimodal than 
a single-mass Gaussian function would predict, and therefore if blue HB stars 
are slightly bluer and red HB stars are slightly redder than a single-mass 
Gaussian would function predict, the fitting, shown in Figure 5, would be 
better.
   One can achieve such near-perfect fits if a bimodal HB mass distribution,
with two mass peaks, is adopted (\cite{dor95}).
   If a single-mass Gaussian function turns out to be a bad approximation,
such efforts with multiple variables would be unavoidable.
   However, one has to be careful in introducing one more parameter in the
already-complex EPS technique.

   Alternatively, the current stellar evolution theory may not be accurate
enough to match the data more closely than what we have already achieved.
   For instance, current models do not take into account any mass loss after
stars arrive at the zero-age HB.
   In fact, low-mass HB stars are good places to expect some mass loss even 
if their surface gravities are higher than those of red giants. 
   This is because the hydrogen-burning shell is so close to the atmosphere 
in those stars.
   Because of their proximity to the energy generating shell, the envelope 
and atmosphere of such hot stars directly feel the impact of the 
hydrogen shell burning that takes place during their HB evolution.
   Demarque \& Eder (1985) showed that a small mass loss rate on the HB
($\dot{M} = 2.5 \times 10^{-10}$ -- $1 \times 10^{-9}$~\Msun/yr) is
enough to generate hot HB stars and to explain the existence of the field sdB 
stars. 
   More detailed and modern calculations would be necessary.

   How important is binary evolution? Additional mass loss may take
place both before and after the helium core flash and create low-mass HB stars 
in binary evolution (Mengel, Norris, \& Gross 1976; \cite{fusi93}; 
\cite{lsg94}).
   In fact, Green et al. (1997) found evidence that most of these hot stars 
in NGC\,6791 are remnants of binary evolution.
   It is important to understand the impact of binary evolution on stellar 
evolution first and its significance in an environment like gEs.
   Binary scenario may attract more attention because UV upturns are 
found to be more significant in the dense cores of gEs.

   If any of these (or other unknown) possibilities mentioned above occurs, 
the presence of hot stars in the field, in some globular clusters, and in 
NGC\,6791 would be easier to understand.
   Detailed theoretical (both evolutionary and hydrodynamical) modeling of
mass loss as well as detections of hot stars, such as the slow blue phase
stars, are crucial for a better understanding of the UV upturn.

   Because UV population models are sensitive to the assumed HB mass
distribution (see also YDO), modelers should explicitly explain their HB 
prescription. 
   There are many EPS models currently, but not many of them provide 
sufficient information about input parameters and assumptions, making 
comparisons very difficult.
   Dorman et al. (1995) and Park \& Lee (1997) are among the few EPS models 
with sufficient information, so we could clarify the differences between
their models and ours.
   Unfortunately, we do not have enough information about the assumed HB
mass distribution in the work of Bressan et al. (1994) whose approach and
conclusion seem to be closest to ours.
 
\subsection{Abundance Distribution in a Galaxy}

   Our study supports the suggestion of Tantalo et al. (1996) and  Kodama 
(1997) that infall models have advantages of matching the near-UV spectrum 
better than simple (non-infall) composite models.
   While such alternatives in chemical evolution models give better fits
to the data, one should also be aware how sensitive EPS models are to 
slight alterations in the adopted chemical evolution model (particularly
in the UV).  
   It must be significant, however, that single age, single metallicity 
models fit the UV observations better than the simplest chemical evolution 
composite models.  

\section{Summary and Conclusions}

   The models based on the metal-rich HB hypothesis seem to satisfy all the 
empirical constraints related with the UV upturn phenomenon for reasonable 
input parameters. 
   Under the conventional assumptions of stellar evolution theory, evolved
low-mass, metal-rich ($Z \gtrsim$~\Zsun) core helium-burning stars are likely 
to be the dominant UV source in gEs, as  suggested by earlier studies
(\cite{gr90}; \cite{bcf94}; \cite{dor95}; \cite{yado95}; \cite{bfdd97}; YDO). 

   The UV upturn is an intricate phenomenon played by an orchestra of various 
instruments among which the following two phenomena have the most profound 
effects.
(1) More metal-rich red giants experience higher mass loss even for a 
$fixed$ mass loss efficiency parameter $\eta$, according to Reimers' formula. 
   This is because a more metal-rich red giant has a higher opacity in the 
atmosphere, and the higher opacity causes a larger stellar radius, and a 
smaller surface gravity, which results in a larger mass loss (see discussion
in Horch et al. (1992) and references therein).
   After the large mass loss, a more metal-rich red giant becomes a 
lower-mass HB star.
   However, a large opacity in metal-rich stars causes even low-mass HB stars
to become red (low \Teff). So mass loss cannot reproduce by itself the
observed magnitude of the UV upturn.
(2) The SBP (slow blue phase) phenomenon, the UV bright phase of the core 
helium-burning stars, is more prominent in more metal-rich stars when a 
positive \DYDZ is assumed.
   Consequently, the magnitude of the UV upturn increases with increasing 
metallicity under the assumption of a positive \DYDZ, which is consistent 
with the empirical discovery (\cite{f83}; \cite{b88}).
   Composite models seem to reproduce the observed range of the magnitude and 
the characteristic temperature of the UV upturn (\Tuv) reasonably.
   However, their match is not as good as those of single abundance models.
   This may indicate a large uncertainty in galactic chemical evolution models.

   If such small mismatchs are considered acceptable, we may claim that
a simple instantaneous burst model of the formation of gEs naturally 
develops a UV upturn with the observed characteristics. 
   The full range of the observed magnitude of the UV upturn (or, \fnuv)
can be produced either by a metallicity dispersion 
(e.g., the majority of stars in UV-strong galaxies are $Z \approx$ 2~\Zsun and
those in UV-weak galaxies are $Z \approx$ \Zsun) or by an age dispersion 
(UV-strong galaxies being older than UV-weak galaxies) among gEs, or 
perhaps both combined.
   At least some age difference seems to be favored if a factor of two 
difference in metallicity among gEs is unlikely.
   Models (single abundance models) also match the observed range of \Tuv 
(the characteristic temperature of the UV upturn) rather precisely. 
   It is interesting to note the apparently reverse correlation between
the magnitude of the UV upturn and \Tuv (with the exception of NGC\,1399), 
although it is not yet statistically significant.
   Under the current assumption of input parameters, this phenomenon is
also understandable by the same scenario that explains the UV 
upturn-metallicity relation.

   The UV upturn may serve as a relative-age indicator, provided the mean 
metallicity in a gE can be independently determined either through 
spectroscopic or photometric studies. 
   However, it seems premature to use the UV upturn as an absolute-age 
indicator until input parameters are much better constrained.
   A more secure calibration of the UV upturn (e.g., for the precise dating 
of galaxies) will require improvements in our understanding of various things
including mass loss during stellar evolution and the origin and evolutionary 
status of sdB stars in gEs.
   Willson et al. (1996) (also \cite{bw91}) have recently claimed that their
hydrodynamical models describe the mass loss better than Reimers' formula with 
a fixed mass loss efficiency parameter.
   It would be a great step forward if such new approaches can provide
the astronomical community with a parameterized mass loss formula for single
stars that matches the observations and is easy to use.
   In the same line, the role and frequency of binary stars, which appear 
responsible for at least a fraction of the sdB's in the stellar population of 
gE's, has to be studied further.  

   More fundamentally, it is crucial to obtain a larger and reliable sample of 
spectra of nearby gEs.
   It is almost unbelievable that there is hardly any gE whose well-calibrated 
spectrum is available from far-UV to infrared.
   It is important to acquire the whole wavelength range of spectrum in order
to find a unique solution using the EPS technique.
   The far-UV (including the $Lyman$ break) spectrum is particularly important 
because it provides important clues to the properties of the UV sources.

   We have not even touched other complexities, such as uncertainties
in the stellar evolutionary calculation, which can be tested best by observing 
nearby stellar populations, and in the spectral library.
   The role of dust and of galaxy merging history could be as important as 
the ones that are discussed here.
   So, the question about the UV upturn is still open.
   However, we feel that the current EPS studies are going toward its solution;
at least, this model (the metal-rich HB hypothesis) has survived so far.

\acknowledgments

   We thank Richard Larson, Wayne Landsman, Tom Brown and Ben Dorman for 
constructive comments. 
   Special thanks to Tom Brown for making the HUT data available and
for many suggestions and clarifications.
   This work was a part of the Ph.D. study of S.Y. (\cite{yi96}) and was
supported in part by NASA grants NAGW-3563 (S.Y. and A.O.), 
NAG5-1486 and NAG5-2469 (P.D.).
   Part of this work was performed while S.Y. held a National Research
Council-(NASA Goddard Space Flight Center) Research Associateship.

{}

\clearpage

\begin{figure}
\epsscale{0.8}
\plotone{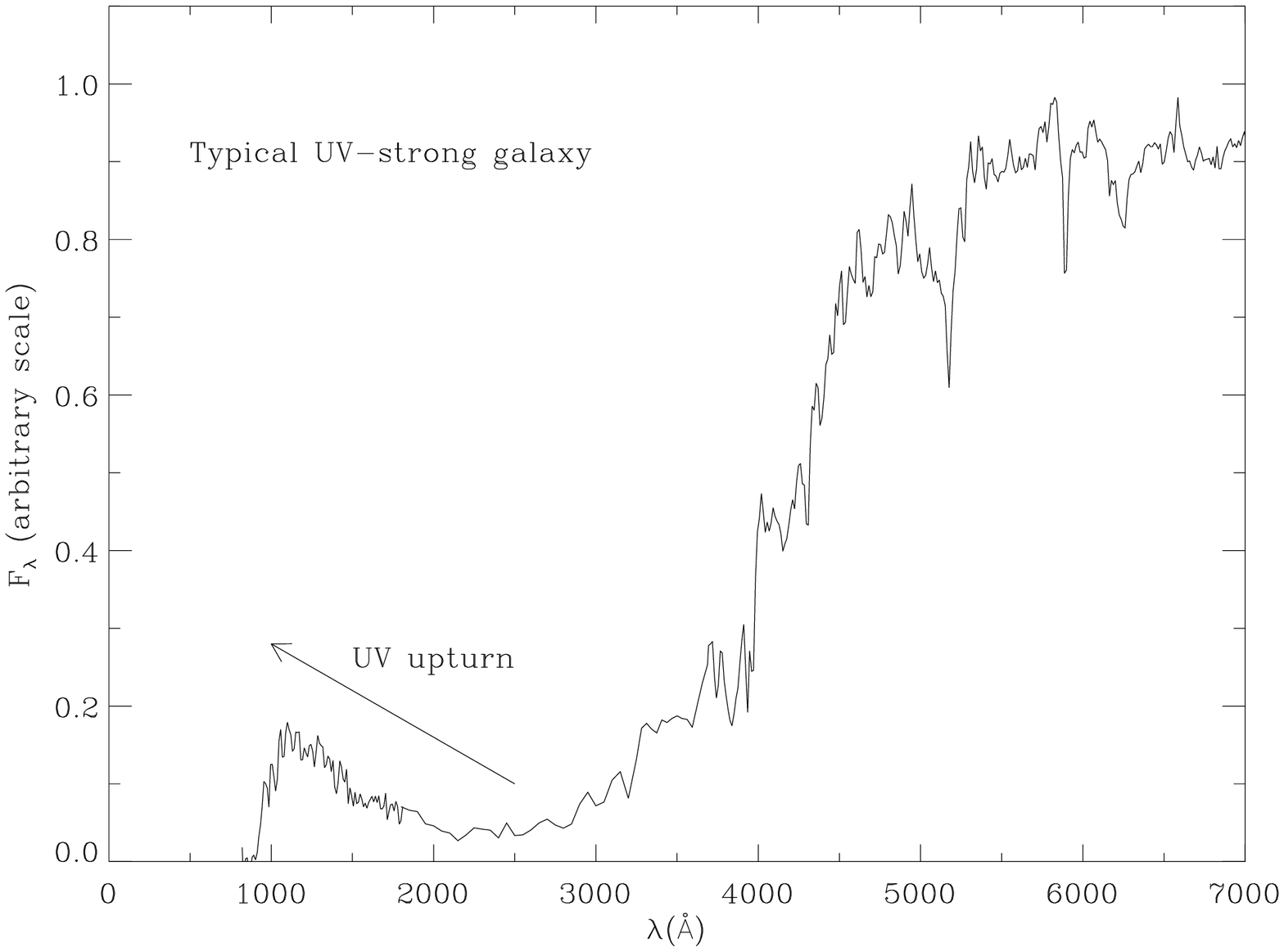}
\caption
{The UV upturn phenomenon in the composite spectrum of a typical UV-strong 
galaxy, NGC\,4552. Sources of the composite spectrum are
(1) $\leq$ 1800 \AA: HUT spectrum of NGC\,4552 (\cite{bfd95}),
(2) 1800 -- 3300 \AA: mean IUE spectrum of UV-strong galaxies
(\cite{b88}), (3) 3300 -- 3700 \AA: UV-strong galaxy NGC\,4649 (\cite{a96}), 
and (4) $\geq$ 3700 \AA:  average of Bica's E1 group galaxies (\cite{bica88}).
Much of this composite spectrum originates from Arimoto (1996).
}
\label{figure 1}
\end{figure}


\begin{figure}
\epsscale{0.8}
\plotone{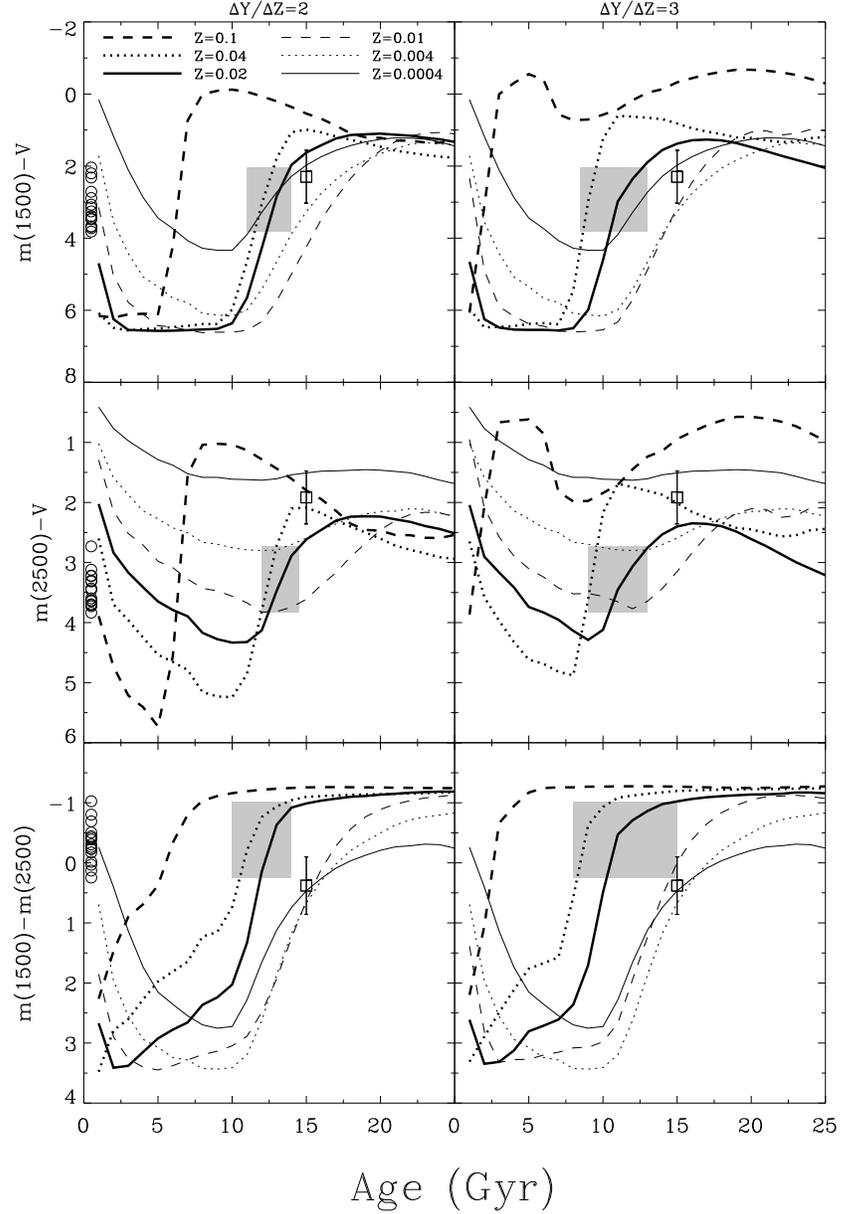}
\caption{UV flux ratios as a function of age and metallicity for single 
abundance models.
The metal-rich ($Z \gtrsim 0.02$) models are for ($\eta$, $x$, $\sigma$) = 
(1.0, 1.35, 0.06), but, $\eta$ = 0.5, and 0.7 have been used for 
$Z = 0.0004$ and 0.004 \& 0.01 models, respectively. 
Observed colors of elliptical galaxies (from Table 2 of Dorman et al. 1995) 
are marked as open circles at the left end of the diagrams, because their
ages are unknown. 
If the majority of stars in gEs are $\approx$ 1 -- 2 \Zsun, the models 
indicate that gEs are 8 -- 15 Gyrs old, as denoted as shaded boxes.
The open boxes with error bars (one s.d.) are the average flux ratios of 
Galactic globular clusters (Table 1 of Dorman et al. 1995) whose average
metallicity is $Z \approx$ 0.0014, assuming the average age of 15 Gyr.
}
\label{Figure 2}
\end{figure}


\begin{figure}
\epsscale{0.8}
\plotone{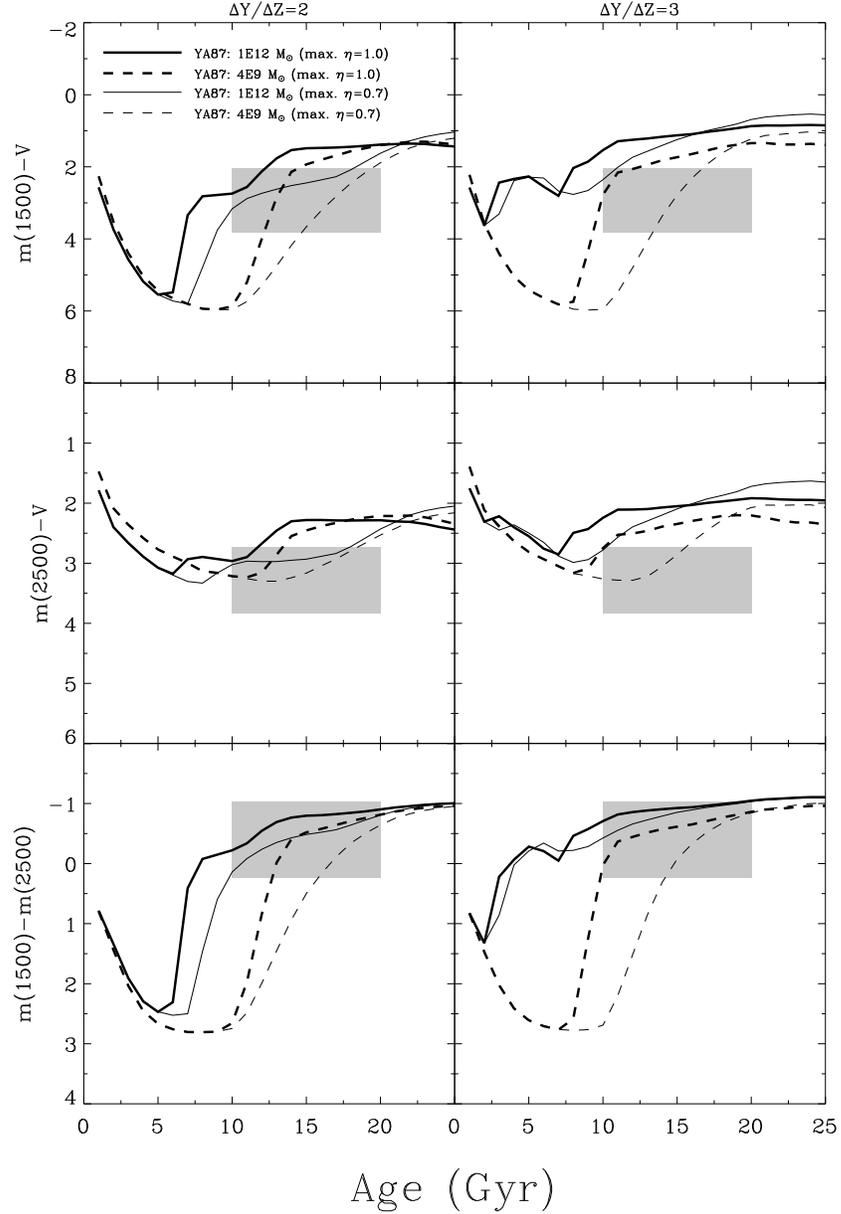}
\caption{Same as Figure 2, but for composite models based on the Yoshii \&
Arimoto abundance distribution models (Yoshii \& Arimoto 1987; YA87).
The thin (thick) lines are based on the variable-$\eta$ models with 
a maximum $\eta$ cutoff at $\eta_{max} = 0.7$ (1.0). See text for details.
The mean abundance of the YA87 $10^{12}$\,\Msun model is approximately 
2~\Zsun ($Z \approx$ 0.04) and that of the $4\,\times\,10^{9}$\,\Msun model
is $\approx$~\Zsun. Note that model \nuvv values are 0.5 -- 1.0 mag bluer than
observed, perhaps indicating that YA87 models predict too many metal-poor 
stars. The shaded boxes are the observed flux ratios assuming that gEs can be
between 10 and 20 Gyrs old.}
\label{Figure 3}
\end{figure}


\begin{figure}
\epsscale{0.8}
\plotone{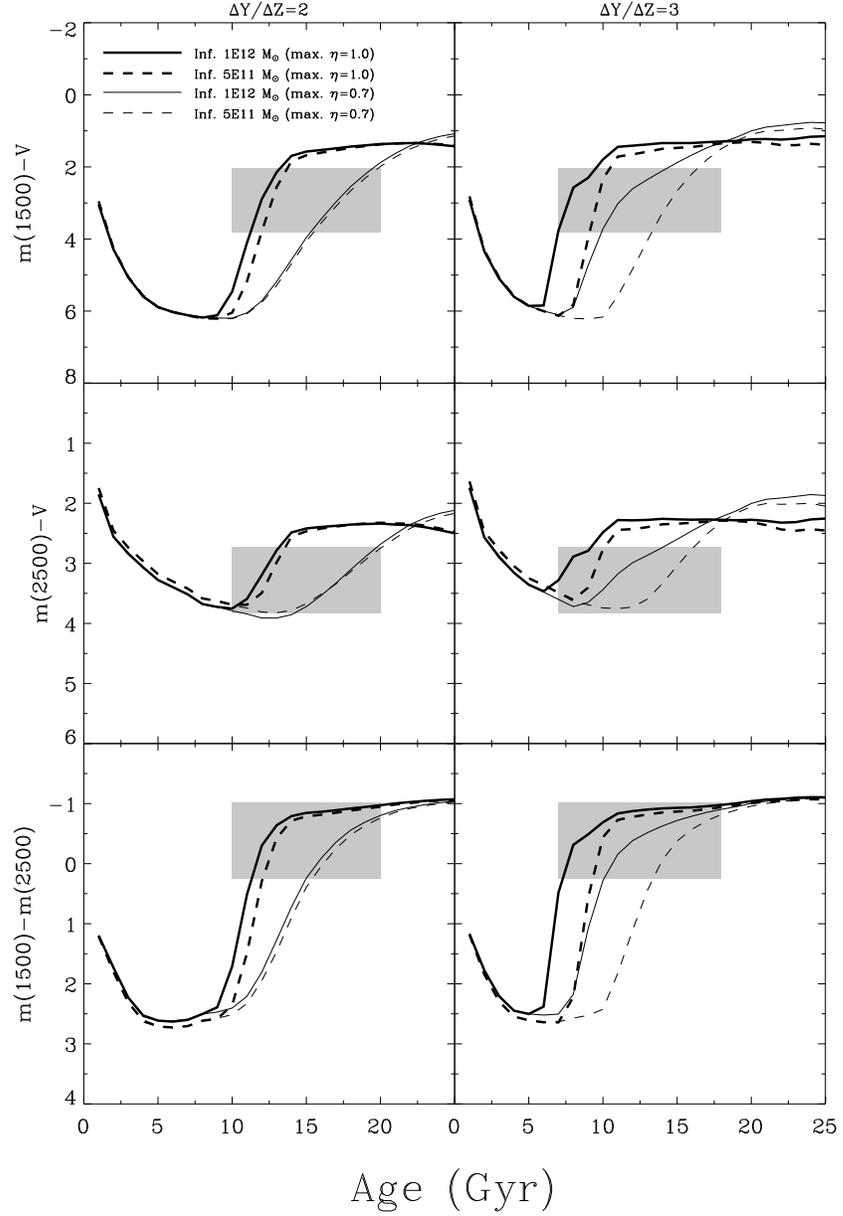}
\caption{Same as Figure 3, but based on the infall model abundance 
distribution of Tantalo et al. (1996).
The mean abundance of their $10^{12}$\,\Msun model is approximately 
2~\Zsun ($Z \approx$ 0.04) and that of the $5\,\times\,10^{10}$\,\Msun model
is somewhat larger than 1 \Zsun. Models are in better agreement with data 
than the models based on the YA87 models do. The shaded boxes indicate the age 
estimates for gEs based on the observed flux ratios observed flux ratios; 
They suggest approximately an age of 10 Gyr (if $\eta_{max}$ = 1.0) --
15 Gyr (if $\eta_{max}$ = 0.7).}
\label{Figure 4}
\end{figure}


\begin{figure}
\epsscale{0.8}
\plotone{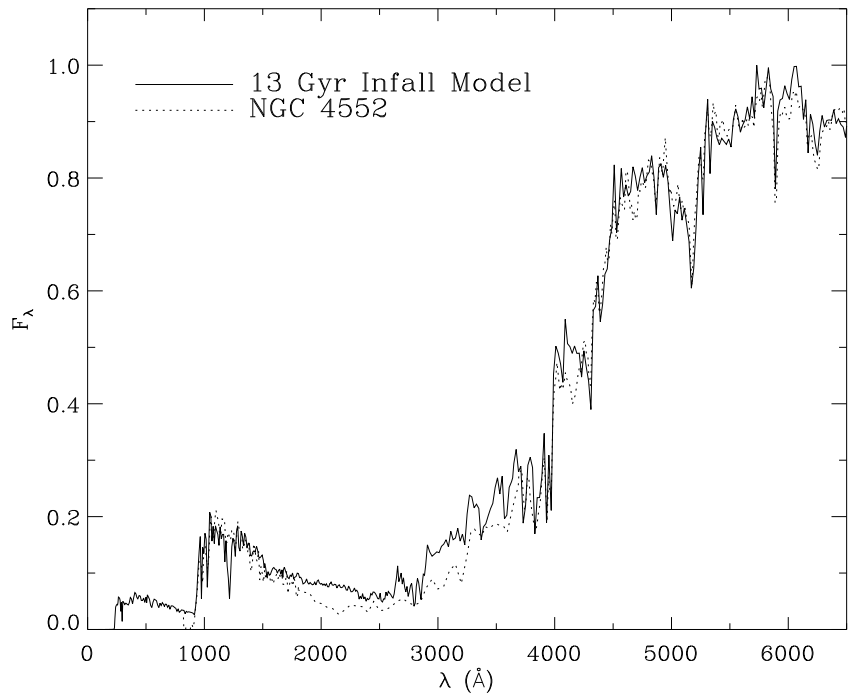}
\caption{Comparison between the empirical spectrum of NGC\,4552 (dotted line)
and the 13 Gyr old composite model (solid line) based on the infall
($10^{12}$\,\Msun ) model. The model fits the data well, but predicts 
too much flux in the near-UV.}
\label{Figure 5}
\end{figure}


\begin{figure}
\epsscale{0.7}
\plotone{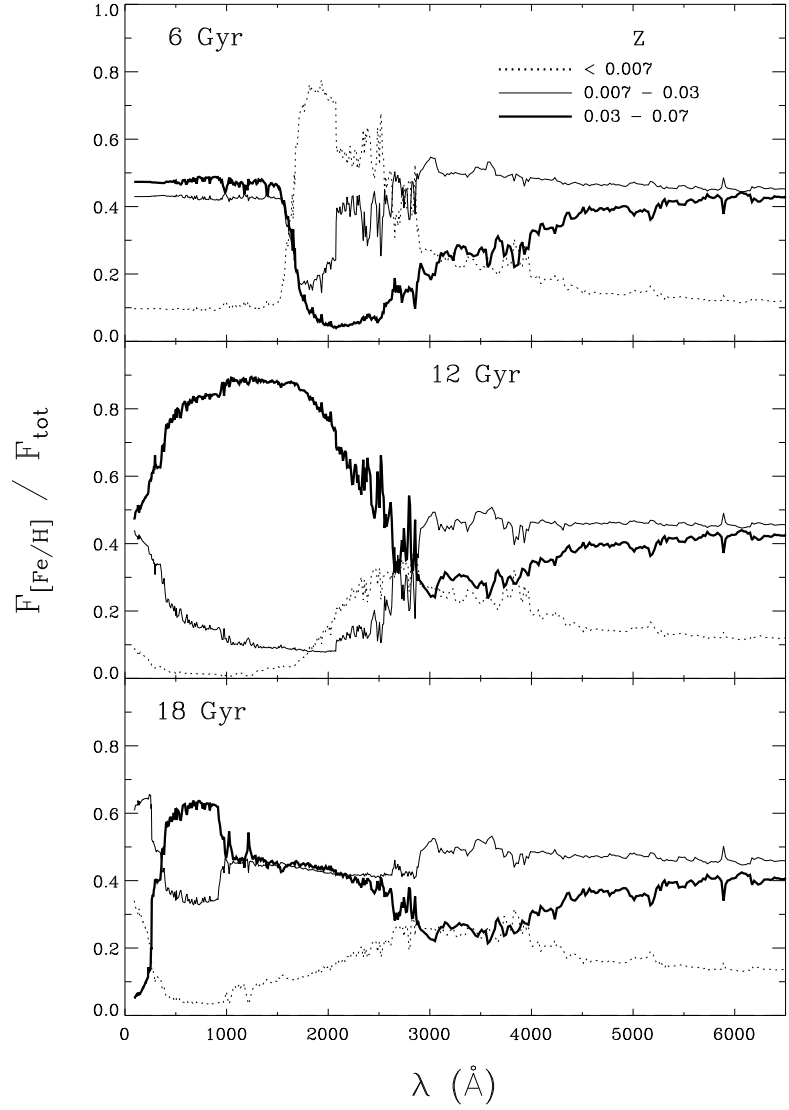}
\caption{Light contribution from various metallicity groups in the infall
models.}
\label{Figure 6}
\end{figure}


\begin{figure}
\epsscale{0.8}
\plotone{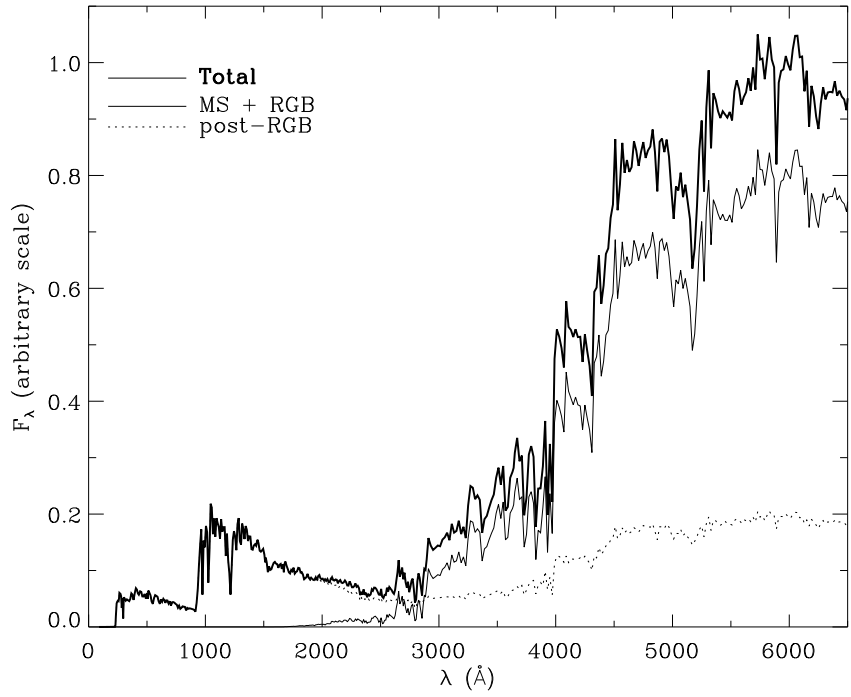}
\caption{Light contribution from different evolutionary stages in the model 
shown in Figure 5.}
\label{figure 7}
\end{figure}


\begin{figure}
\epsscale{0.8}
\plotone{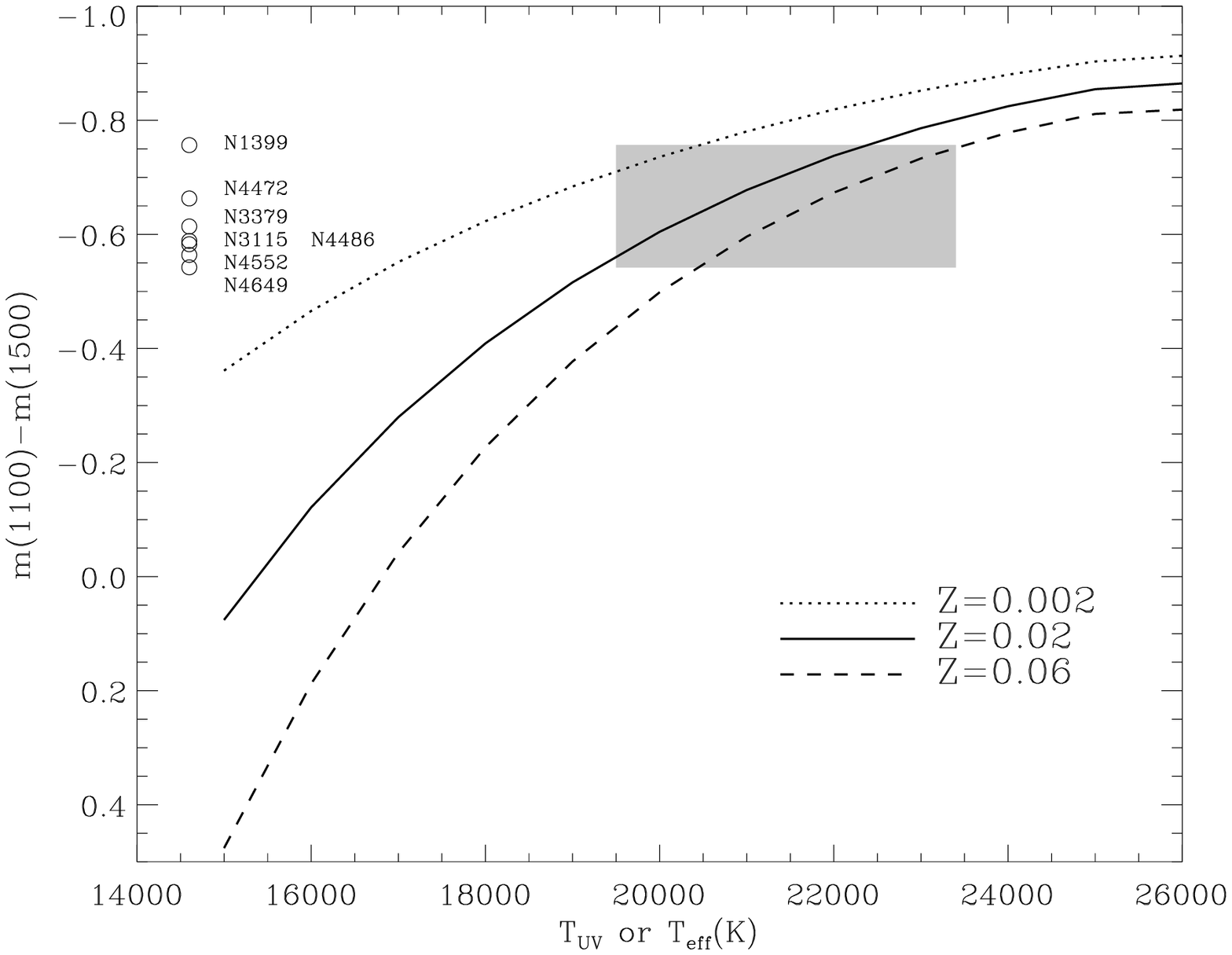}
\caption{The m(1100)-m(1500) of an individual star as a function of \Teff and 
metallicity. The m(1100)-m(1500) has been measured from the Kurucz model 
stellar spectra of \logg = 5.0 (\cite{k92}).
Because m(1100)-m(1500) is sensitive to \Teff, it serves as a \Tuv indicator
for composite systems.
Open circles are the measured colors from the HUT spectra (\cite{bfd95}).
If hot stars in ellipticals are metal-rich ($\gtrsim$ \Zsun), our \Tuv
estimates ($\approx$ 19,500 -- 23,400 K) are in agreement with the HUT 
estimate, 20,000 -- 23,000 K.}
\label{figure 8}
\end{figure}


\begin{figure}
\epsscale{0.8}
\plotone{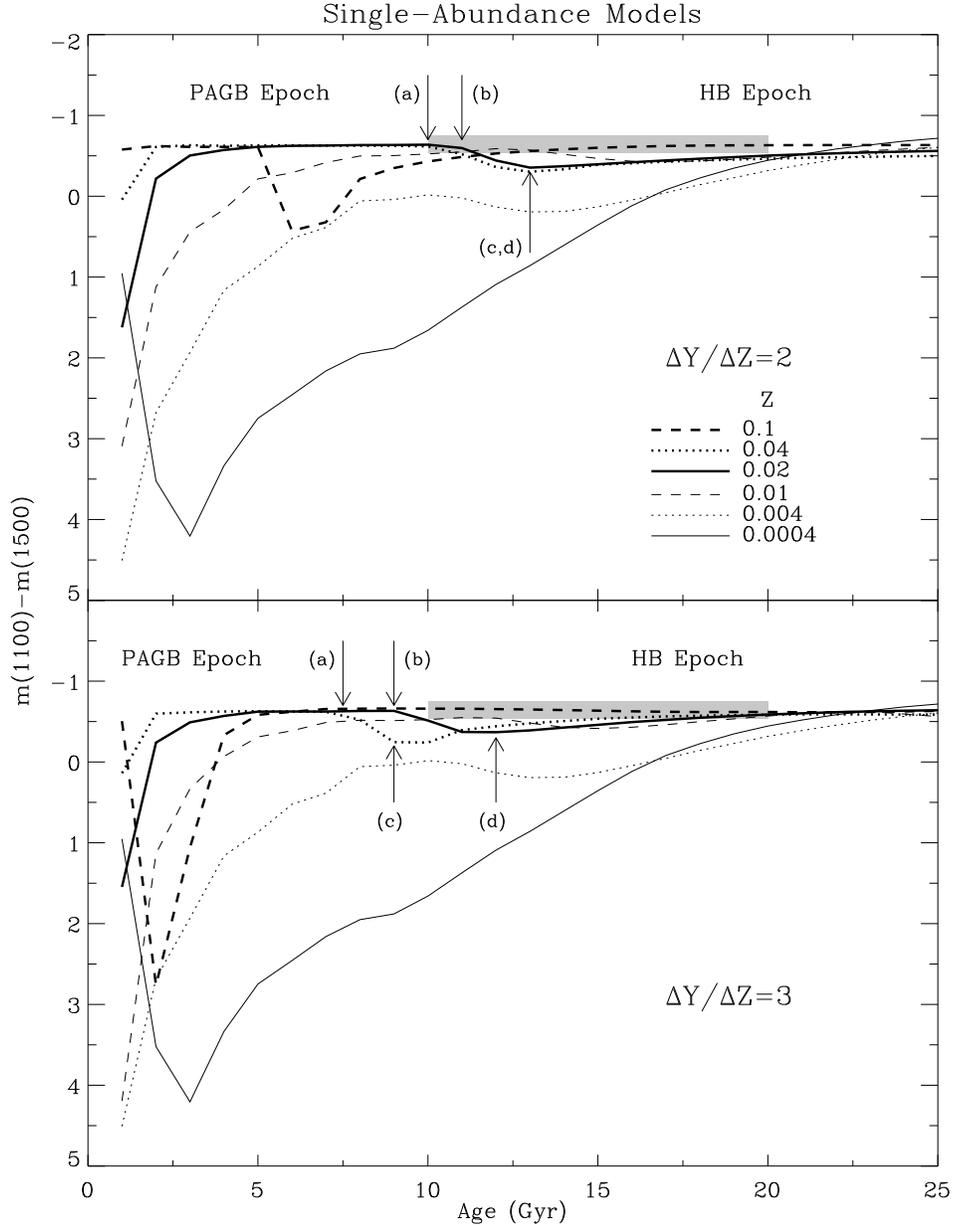}
\caption{An index of the characteristic temperature of the UV upturn (\Tuv), 
m(1100)-m(1500), as a function of age and metallicity for single abundance 
models. See Figure 2 for input parameters. The grey box is the observed range 
of m(1100)-m(1500), assuming that gEs are about 10 -- 20 Gyrs old. 
Models of $Z =$~\Zsun are in agreement with the 
observed values. After turning points (a) \& (b), dominant UV source
changes from PAGB stars to HB stars for the population of $Z$ = 0.04 \& 0.02,
respectively. After (c) \& (d), HB stars completely dominate the UV spectrum.
}
\label{figure 9}
\end{figure}


\begin{figure}
\epsscale{0.8}
\plotone{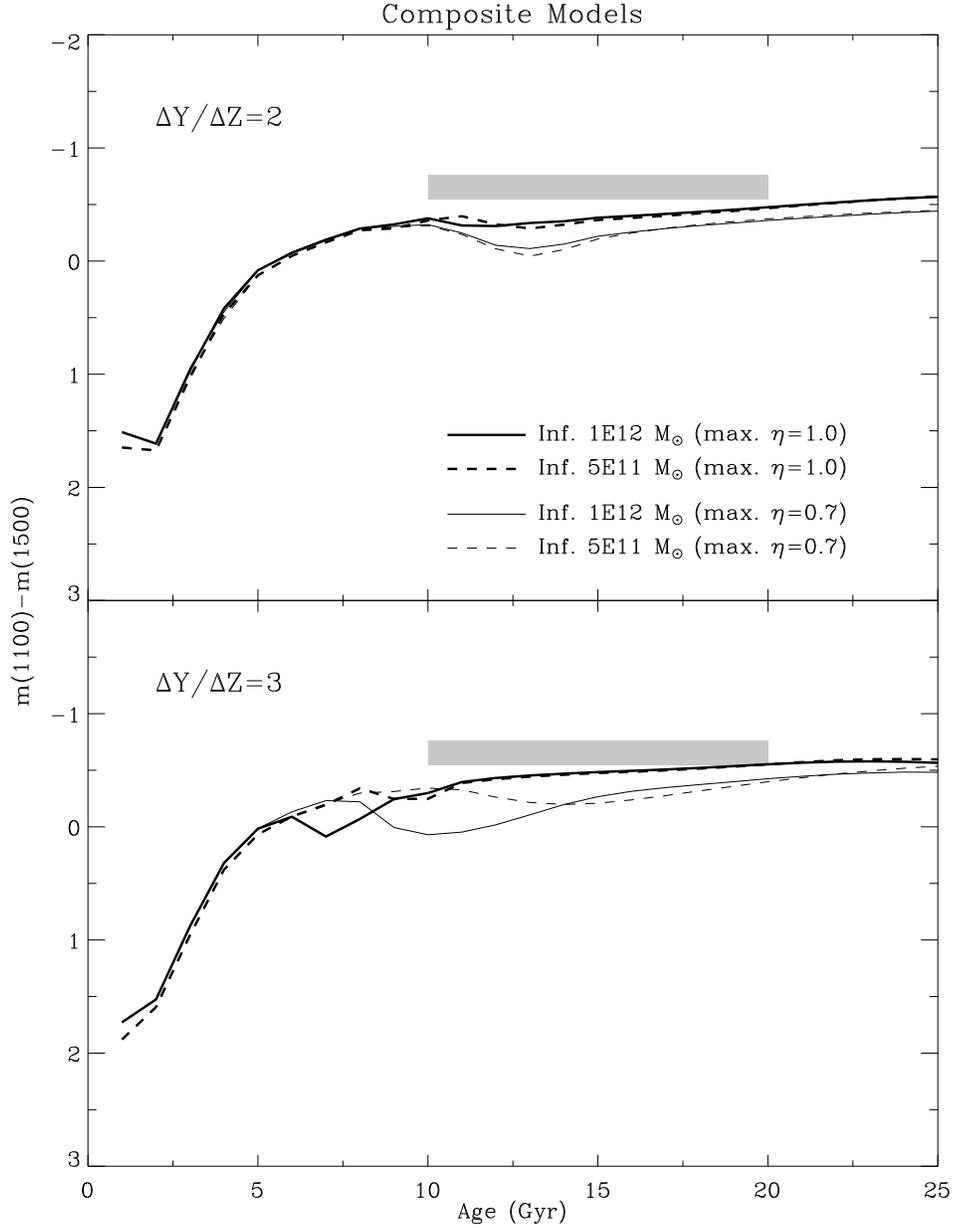}
\caption{Same as Figure 9, but for composite (Infall) models, as described in 
Figure 4. Models are based on the variable-$\eta$ hypothesis with $\eta_{max}$ 
= 1.0 (thick lines) and with $\eta_{max}$ = 0.7 (thin lines), respectively. 
}
\label{figure 10}
\end{figure}


\begin{figure}
\epsscale{0.8}
\plotone{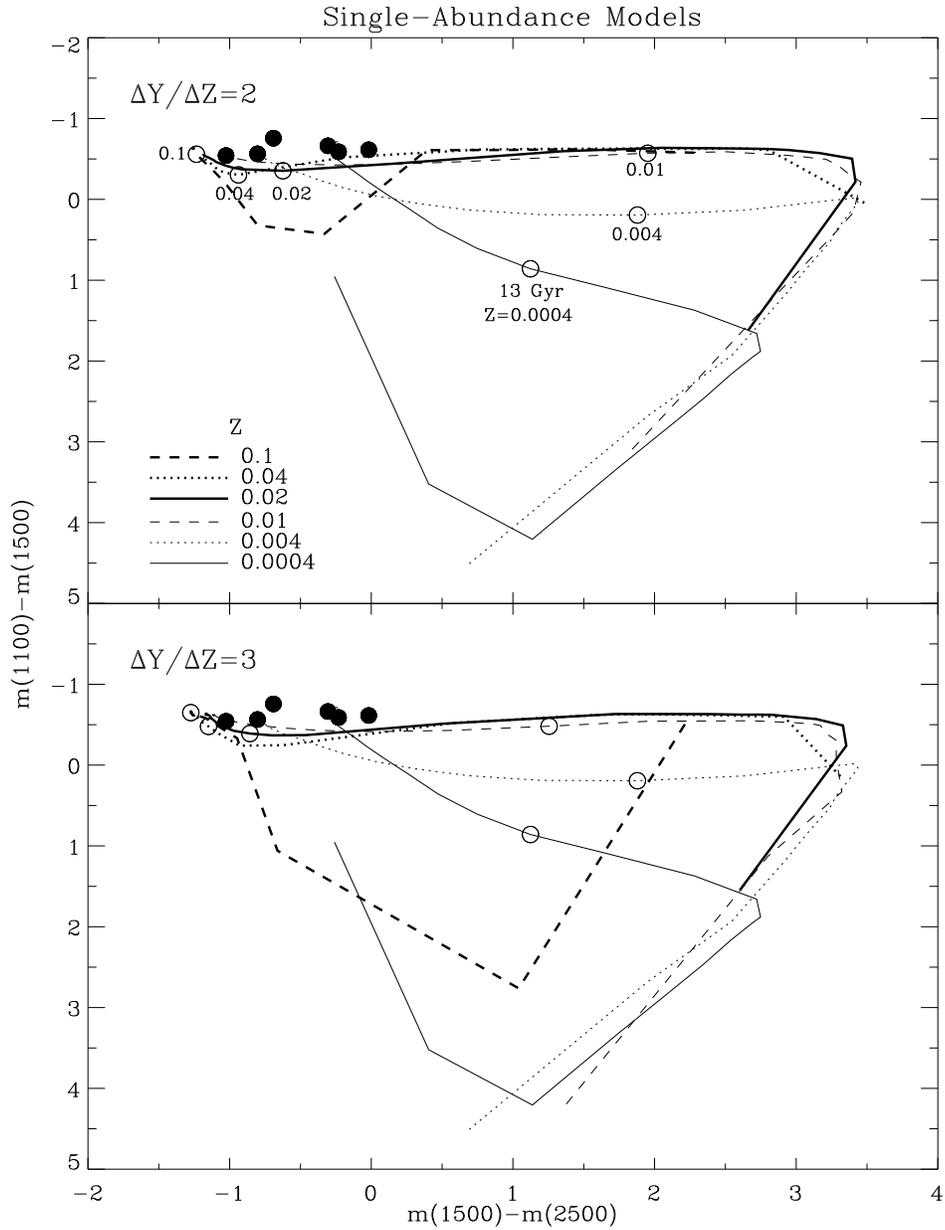}
\caption{Two color diagrams for single abundance models. 
Each model line spans 1 -- 25 Gyr of age, with 13 Gyr models (open circles)
marked to be compared with empirical data (filled circles). The same input 
parameters as in Figure 2 have been used.}
\label{figure 11}
\end{figure}


\begin{figure}
\epsscale{0.8}
\plotone{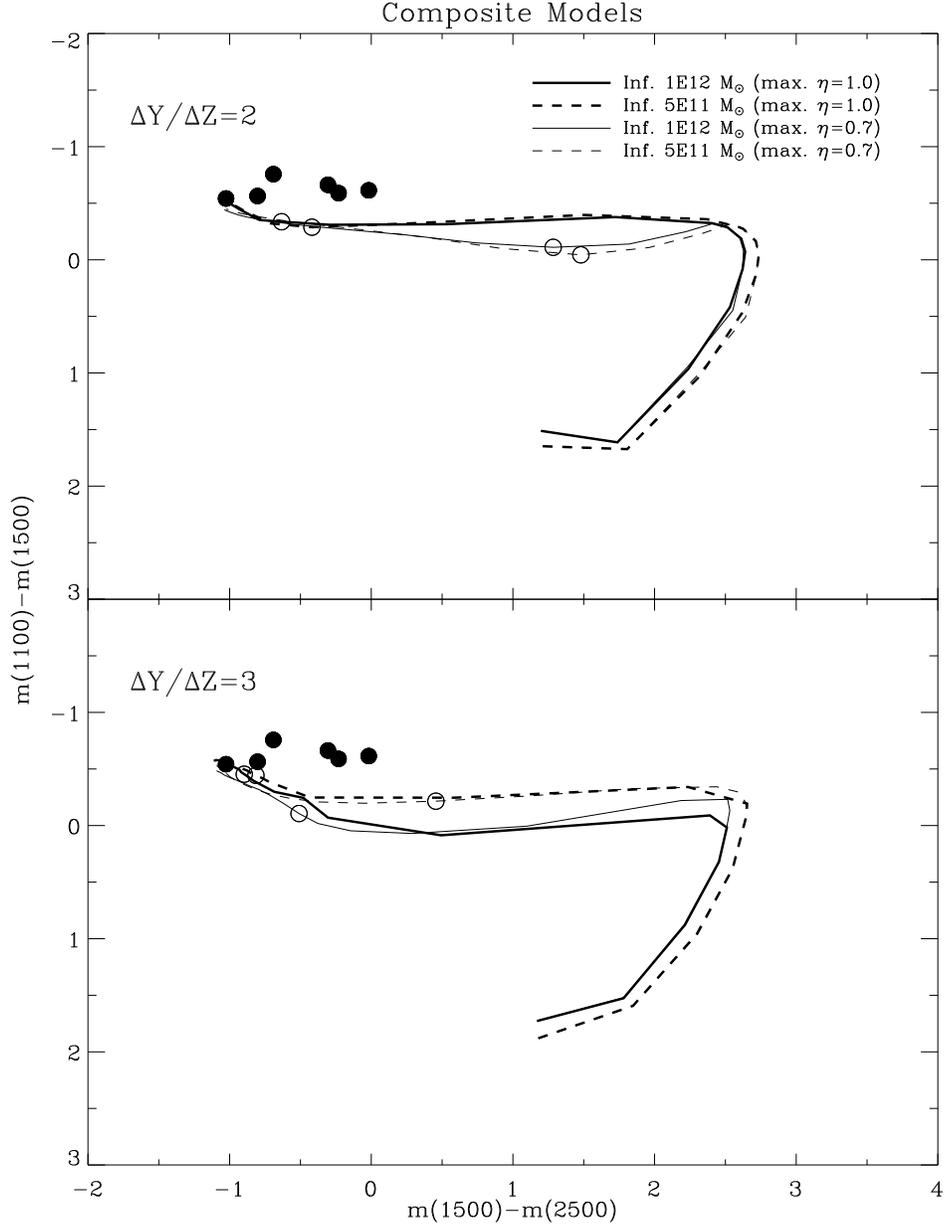}
\caption{Same as figure 11, but for composite (Infall) models.
Models are based on the variable-$\eta$ hypothesis with $\eta_{max}$ = 1.0
(thick lines) and with $\eta_{max}$ = 0.7 (thin lines), respectively. 
}
\label{figure 12}
\end{figure}

\end{document}